\documentclass[twocolumn,tighten,times]{aastex631}

\PassOptionsToPackage{dvipsnames,svgnames,x11names}{xcolor}
\usepackage{tabularx}

\usepackage{amsmath}  
\usepackage{amssymb}  
\usepackage[T1]{fontenc}
\usepackage{xspace}
\usepackage{multirow}
\usepackage{xcolor}

\defcitealias{kenworthy_salt3_2021}{K21}

\usepackage{newunicodechar,graphicx}
\DeclareRobustCommand{\okina}{%
  \raisebox{\dimexpr\fontcharht\font`A-\height}{%
    \scalebox{0.8}{`}%
  }%
}
\newunicodechar{ʻ}{\okina}

\newif{\ifchangetext}
\changetextfalse

\InputIfFileExists{\jobname-options}

\ifchangetext
  
  \newcommand{\changenote}[1]{\textcolor{blue}{ \bf #1}}x
\else
  
  \newcommand{\changenote}[1]{}
\fi

\hypersetup{
    colorlinks=True,
    citecolor=black,
    linkcolor=black,
    filecolor=magenta,      
    urlcolor=cyan,
}

\newcommand{\angstrom}{\mbox{\normalfont\AA}\xspace}

\def\arcsec{\ensuremath{^{\prime\prime}}}

\def\xone{\ensuremath{x_{1}}}
\def\C{\ensuremath{\mathcal{C}}}

\newcommand{\STScI}{Space Telescope Science Institute, Baltimore, MD 21218, USA}
\newcommand{\JHU}{Physics and Astronomy Department, Johns Hopkins University, Baltimore, MD 21218, USA}

\newcommand{\ISEF}{ISEF International Fellowship}
\newcommand{\NEF}{NASA Einstein Fellow}
\newcommand{\snz}{2.15}
\newcommand{\snzerr}{0.01}
\newcommand{\mufit}{46.10}
\newcommand{\mufiterrl}{0.18}
\newcommand{\mufiterrh}{0.17}

\usepackage{xspace}
\usepackage{changepage}

\newcommand{\highzIa}{SN\,$2023$aeax\xspace}
\usepackage[mathlines]{lineno}

\begin{document}
    
    \title{\vspace{-35pt}Testing for Intrinsic Type Ia Supernova Luminosity Evolution at $\mathbf{z>2}$ with \textit{JWST}\vspace{-8pt}}

\author[0000-0002-2361-7201
]{J.~D.~R.~Pierel}
\correspondingauthor{J.~D.~R.~Pierel} 
\email{jpierel@stsci.edu}
\altaffiliation{\NEF}
\affiliation{\STScI}

\author[0000-0003-4263-2228]{D.~A.~Coulter} 
\affiliation{\STScI} 

\author[0000-0003-2445-3891]{M.~R.~Siebert} 
\affiliation{\STScI}

\author[0000-0003-3596-8794]{H.~B.~Akins}
\affiliation{The University of Texas at Austin, 2515 Speedway Blvd Stop C1400, Austin, TX 78712, USA}

\author[0000-0003-0209-674X]{M.~Engesser}
\affiliation{\STScI}

\author[0000-0003-2238-1572]{O.~D.~Fox} 
\affiliation{\STScI}

\author[0000-0002-3560-8599]{M.~Franco}
\affiliation{CEA, Université Paris-Saclay, Université Paris Cité, CNRS, AIM, 91191, Gif-sur-Yvette, France}
\affiliation{The University of Texas at Austin, 2515 Speedway Blvd Stop C1400, Austin, TX 78712, USA}

\author[0000-0002-4410-5387]{A.~Rest} 
\affiliation{\STScI}
\affiliation{\JHU}

\author[0009-0008-1965-9012]{A.~Agrawal}
\affiliation{Department of Astronomy, University of Illinois at Urbana-Champaign, 1002 W. Green St., IL 61801, USA}

\author[0009-0007-8764-9062]{Y.~Ajay}
\affiliation{\JHU}

\author[0000-0001-9610-7950]{N.~Allen}
\affiliation{Cosmic Dawn Center (DAWN), Denmark} 
\affiliation{Niels Bohr Institute, University of Copenhagen, Jagtvej 128, DK-2200, Copenhagen, Denmark}

\author[0000-0002-0930-6466]{C.~M.~Casey}
\affiliation{Department
 of Physics, University of California, Santa Barbara, Santa Barbara, CA 93109, USA}

\affiliation{The
 University of Texas at Austin, 2515 Speedway Blvd Stop C1400, Austin, TX 78712, USA}
\affiliation{Cosmic
 Dawn Center (DAWN), Denmark}

\author[0000-0002-4781-9078]{C.~DeCoursey}
\affiliation{Steward Observatory, University of Arizona, 933 N. Cherry Avenue, Tucson, AZ 85721 USA}

\author[0000-0003-4761-2197]{N.~E.~Drakos}
\affiliation{Department of Physics and Astronomy, University of Hawaii, Hilo, 200 W Kawili St, Hilo, HI 96720, USA}

\author[0000-0003-1344-9475]{E.~Egami}
\affiliation{Steward Observatory, University of Arizona, 933 N. Cherry Avenue, Tucson, AZ 85721 USA}

\author[0000-0002-9382-9832]{A.~L.~Faisst}
\affiliation{Caltech/IPAC, MS 314-6, 1200 E. California Blvd. Pasadena, CA 91125, USA}

\author[0000-0003-3703-5154]{S.~Gezari}
\affiliation{\STScI}

\author[0000-0002-0236-919X]{G.~Gozaliasl}
\affiliation{Department of Computer Science, Aalto University, P.O. Box 15400, FI-00076 Espoo, Finland}
\affiliation{Department of Physics, University of, P.O. Box 64, FI-00014 Helsinki, Finland}

\author[0000-0002-7303-4397]{O.~Ilbert}
\affiliation{Aix Marseille Univ, CNRS, CNES, LAM, Marseille, France  }

\author[0000-0002-6230-0151]{D.~O.~Jones} 
\affiliation{Institute for Astronomy, University of Hawaiʻi, 640 N. A’ohoku Pl., Hilo, HI 96720, USA}

\author[0000-0003-2495-8670]{M.~Karmen} 
\affiliation{\JHU}

\author[0000-0001-9187-3605]{J.~S.~Kartaltepe}
\affiliation{Lab for Multiwavelength Astrophysics, School of Physics and Astronomy, Rochester Institute of Technology, 84 Lomb Memorial Dr., Rochester, NY 14623, USA}

\author[0000-0002-6610-2048]{A.~M.~Koekemoer}
\affiliation{\STScI}

\author[0009-0003-8380-4003]{Z.~G.~Lane}
\affiliation{School of Physical and Chemical Sciences — Te Kura Matū, University of Canterbury, Private Bag 4800, Christchurch 8140, New Zealand}

\author[0000-0003-2366-8858]{R.~L.~Larson}
\affiliation{Lab for Multiwavelength Astrophysics, School of Physics and Astronomy, Rochester Institute of Technology, 84 Lomb Memorial Dr., Rochester, NY 14623, USA}

\author[0009-0005-5008-0381]{T.~Li} 
\affiliation{Institute of Cosmology and Gravitation, University of Portsmouth, Burnaby Road, Portsmouth, PO1 3FX, UK
}

\author[0000-0001-9773-7479]{D.~Liu}
\affiliation{Purple Mountain Observatory, Chinese Academy of Sciences, 10 Yuanhua Road, Nanjing 210023, China}

\author[0000-0003-1169-1954]{T.~J.~Moriya}
\affiliation{National Astronomical Observatory of Japan, National Institutes of Natural Sciences, 2-21-1 Osawa, Mitaka, Tokyo 181-8588, Japan}
\affiliation{Graduate Institute for Advanced Studies, SOKENDAI, 2-21-1 Osawa, Mitaka, Tokyo 181-8588, Japan}
\affiliation{School of Physics and Astronomy, Monash University, Clayton, Victoria 3800, Australia}

\author[0000-0002-9489-7765]{H.~J.~McCracken}
\affiliation{Institut d’Astrophysique de Paris, UMR 7095, CNRS, and Sorbonne Université, 98 bis boulevard Arago, F-75014 Paris, France}

\author[0000-0003-2397-0360]{L.~Paquereau} 
\affiliation{Institut d’Astrophysique de Paris, UMR 7095, CNRS, and Sorbonne Université, 98 bis boulevard Arago, F-75014 Paris, France}

\author[0000-0001-9171-5236]{R.~M.~Quimby}
\affiliation{Department of Astronomy/Mount Laguna Observatory, SDSU, 5500 Campanile Drive, San Diego, CA 92812-1221, USA}
\affiliation{Kavli Institute for Physics and Mathematics of the Universe, The U. of Tokyo, Kashiwa, Chiba 277-8583, Japan}

\author[0000-0003-0427-8387]{R.~M.~Rich}
\affiliation{Department of Physics and Astronomy, UCLA, PAB 430 Portola Plaza, Box 951547, Los Angeles, CA 90095-1547}

\author[0000-0002-4485-8549]{J.~Rhodes}
\affiliation{Jet Propulsion Laboratory, California Institute of Technology, 4800 Oak Grove Drive, Pasadena, CA 91001, USA}

\author[0000-0002-4271-0364]{B.~E.~Robertson}
\affiliation{Department of Astronomy and Astrophysics, University of California, Santa Cruz, 1156 High Street, Santa Cruz, CA 95064, USA}

\author[0000-0002-1233-9998]{D.~B.~Sanders}
\affiliation{Institute for Astronomy, University of Hawai’i at Manoa, 2680 Woodlawn Drive, Honolulu, HI 96822, USA}

\author[0000-0002-9301-5302]{M.~Shahbandeh} 
\affiliation{\STScI}

\author[0000-0002-7087-0701]{M.~Shuntov}
\affiliation{Cosmic Dawn Center (DAWN), Denmark} 
\affiliation{Niels Bohr Institute, University of Copenhagen, Jagtvej 128, DK-2200, Copenhagen, Denmark}

\author[0000-0002-0000-6977]{J.~D.~Silverman}
\affiliation{Kavli Institute for Physics and Mathematics of the Universe, The U. of Tokyo, Kashiwa, Chiba 277-8583, Japan}
\affiliation{Department of Astronomy, School of Science, The University of Tokyo, 7-3-1 Hongo, Bunkyo, Tokyo 113-0033, Japan}

\author[0000-0002-7756-4440]{L.~G.~Strolger} 
\affiliation{\STScI}

\author[0000-0003-3631-7176]{S.~Toft}
\affiliation{Cosmic Dawn Center (DAWN), Denmark} 
\affiliation{Niels Bohr Institute, University of Copenhagen, Jagtvej 128, DK-2200, Copenhagen, Denmark}

\author[0000-0002-0632-8897]{Y.~Zenati}
\altaffiliation{\ISEF}
\affiliation{\JHU}
\affiliation{\STScI}


\begin{abstract}

The \textit{James Webb Space Telescope} (\textit{JWST}) is opening new frontiers of transient discovery and follow-up at high-redshift. Here we present the discovery of a spectroscopically confirmed Type Ia supernova (SN\,Ia; \highzIa) at $z=\snz$ with \textit{JWST}, including a NIRCam multi-band light curve. \highzIa lands at the edge of traditional low-$z$ cosmology cuts because of its blue color (peak rest-frame $B-V\sim-0.3$) but with a normal decline rate ($\Delta m_{15}(B)\sim1.25$), and applying a fiducial standardization with the BayeSN model we find the \highzIa luminosity distance is in $\sim0.1\sigma$ agreement with $\Lambda$CDM. \highzIa is only the second spectroscopically confirmed SN\,Ia in the dark matter-dominated Universe at $z>2$ (the other is SN\,$2023$adsy), giving it rare leverage to constrain any potential evolution in SN\,Ia standardized luminosities. Similar to SN\,$2023$adsy ($B-V\sim0.8)$, \highzIa has a fairly extreme (but opposite) color, which may be due to the small sample size or a secondary factor, such as host galaxy properties. Nevertheless, the \highzIa spectrum is well-represented by normal low-$z$ SN\,Ia spectra and we find no definitive evolution in SN\,Ia standardization with redshift. Still, the first two spectroscopically confirmed $z>2$ SNe\,Ia have peculiar colors and combine for a $\sim1\sigma$ distance slope relative to $\Lambda$CDM, though in agreement with recent SN\,Ia cosmological measurements.

\end{abstract}

\section{Introduction}
\label{sec:intro}
Luminosity distances measured from Type Ia supernovae (SNe\,Ia) have led to the inferred existence of dark energy \citep{riess_observational_1998,perlmutter_measurements_1999}, our most precise local measurement of the Hubble constant \citep[$H_0$][]{riess_comprehensive_2022}, and have constrained numerous astrophysical quantities (e.g, the rate of light and intermediate elements, nucleosynthesis, etc). SNe\,Ia will continue to refine our understanding of dark energy through upcoming missions such as the Rubin Observatory Legacy Survey of Space and Time (LSST) and \textit{Roman Space Telescope} High Latitude Time Domain Survey (HLTDS), which will use large numbers of SNe\,Ia discovered over a wide redshift range \citep[$z\lesssim3$;][]{hounsell_simulations_2018,ivezic_lsst_2019,rose_reference_2021,mitra_using_2023} to measure changes in dark energy over time. 

Current measurements of evolving dark energy require that the standardization properties of SNe\,Ia do not change with redshift, as such a signal could falsely mimic a change in dark energy. However, SN\,Ia luminosity evolution is a distinct possibility as many redshift-evolving global properties could plausibly impact SN\,Ia luminosities \citep[e.g., metallicity, dust;][]{moreno-raya_dependence_2016,brout_its_2021}. As dark energy is not expected to vary significantly in the dark-matter dominated universe beyond $z\sim2$, high-$z$ SNe\,Ia have unique leverage on these potential SN\,Ia systematics \citep{riess_first_2006,lu_constraints_2022}. If luminosity distances measured from SNe\,Ia at $z\gtrsim2$ diverge from expectations \citep[i.e., existing cosmological measurements;][]{scolnic_complete_2018,brout_pantheon_2022,des_collaboration_dark_2024}, it would strongly indicate the existence of intrinsic SN\,Ia luminosity evolution \citep[for example, see][]{moreno-raya_dependence_2016} or highly non-standard cosmological model, instead of variable dark energy.

Detecting SNe\,Ia at $z\gtrsim2$ requires deep (m$_{AB}\gtrsim26$ per-visit depth) imaging observations in red ($\gtrsim1.2\mu$m) filters. While the \textit{Hubble Space Telescope} (\textit{HST}) observed SNe\,Ia to $z=2.24$, all SNe\,Ia beyond $z\sim1.6$ were strongly lensed \citep[which adds many systematics, see][]{pierel_projected_2021,chen_jwst_2024,frye_jwst_2024,pierel_jwst_2024,pierel_lensed_2024,pascale_sn_2025} and/or photometrically classified \citep{rodney_type_2014}. Due to both depth and wavelength constraints \citep[e.g.,][]{filippenko_optical_1997}, spectroscopic confirmation of SNe\,Ia at $z>2$ was simply not possible before the arrival of the \textit{James Webb Space Telescope} (\textit{JWST}). Since its launch, the sensitivity and wavelength coverage of \textit{JWST} has allowed it to consistently make detections and spectroscopic classifications of high-$z$ SNe   \citep[$z\lesssim4$;][]{engesser_detection_2022,engesser_discovery_2022, decoursey_discovery_2023,decoursey_jades_2024,pierel_jwst_2024, pierel_lensed_2024,pierel_discovery_2024,siebert_discovery_2024,coulter_discovery_2025}, revolutionizing our view of the high-$z$ transient universe.  

The first spectroscopically confirmed SN\,Ia at $z>2$, SN\,$2023$adsy, was recently discovered and followed with \textit{JWST} \citep{decoursey_jades_2024,pierel_discovery_2024}. SN\,$2023$adsy was classified both spectroscopically and photometrically as a SN\,Ia, with a spectroscopic redshift of $z=2.903$, but with some peculiarties. In particular, the SN was extremely red (rest-frame color $B-V\sim0.8$, compared to normal low-$z$ SNe\,Ia $-0.4\lesssim B-V\lesssim0.2$) with a high Ca\,II velocity $\sim19,000$\,km/s relative to the normal population of SNe\,Ia. Spectral energy distribution (SED) modeling of the SN host template photometry show the galaxy is fairly low-mass ($\sim10^8 M_\odot$), low-metallicity ($\sim0.3 Z_\odot$), and low-extinction
($A_V < 0.1$), suggesting the SN may be intrinsically red. Using traditional standardization methods \citep[e.g.,][]{guy_salt2:_2007,betoule_improved_2014,scolnic_complete_2018,kenworthy_salt3_2021,pierel_salt3nir_2022} resulted in a SN\,$2023$adsy luminosity distance $\sim1\sigma$ above $\Lambda$CDM \citep{pierel_discovery_2024}. Thus the properties of SN\,$2023$adsy suggest that SN\,Ia population characteristics may be changing with redshift \citep[i.e., changes in the progenitor system may lead to systematically redder, Ca-rich SNe\,Ia, see][]{woosley_models_1986,bildsten_faint_2007,perets_faint_2010,shen_thermonuclear_2010,waldman_helium_2011,kasliwal_calcium-rich_2012,foley_kinematics_2015,de_zwicky_2020,zenati_origins_2023}, and the stability of their standardization is still uncertain to $z\sim3$. SN\,$2023$adsy is the first spectroscopically confirmed SN\,Ia at $z>2$, and thus a larger sample is needed to determine if SN\,$2023$adsy is representative or peculiar for high-$z$ SNe\,Ia and to conclusively determine if their standardization evolves with redshift. 

A candidate for the second confirmed non-strongly lensed SN\,Ia at $z>2$ has been found in \textit{JWST} imaging conducted as part of the \textit{JWST} Cosmic Evolution Survey \citep[COSMOS-Web;][]{casey_cosmos-web_2023} program. COSMOS-Web observed $0.54\deg^2$ of sky to single-visit $5\sigma$ depths of $m_{AB}\sim28$ in $4$ NIRCam filters in January 2024, overlapping with $148\arcmin^2$ of imaging (containing $8$ NIRCam filters) from the Public Release IMaging for Extragalactic Research \citep[PRIMER;][]{dunlop_primer_2021} program taken in Spring 2023. This repeated imaging gave a sufficiently long baseline to search for transient objects with sensitivity for SNe\,Ia to $z>3$. Of the dozens of detected transient objects (A.~Rest et al., in preparation), one (subsequently named \highzIa and found in a galaxy at R.A.$=10$h$00$m$26.3911$s decl.$=+02$d$19$m$34.077$s), was identified by its colors, photometric redshift, and luminosity as a possible SN\,Ia candidate at $z\sim2.3$. A \textit{JWST} Director's Discretionary Time (DDT) program was approved to follow-up the most interesting transients in the field \citep{coulter_high-z_2024}, providing an additional imaging epoch and a spectrum for several SNe including \highzIa, which received a refined redshift of $z=\snz\pm\snzerr$ based on the spectrum.

Here we describe the classification and analysis of \highzIa, beginning with a summary of the observations in Section \ref{sec:obs} and followed by a spectroscopic classification in Section \ref{sec:class}. Light curve fitting and the subsequent standardized distance measurement are completed in Section \ref{sec:distance}, and we conclude in Section \ref{sec:conclusion} with a discussion of the implications of \highzIa for SN\,Ia cosmology and the future of high-$z$ SN\,Ia observations with \textit{JWST}. In this analysis, we assume a standard flat $\Lambda$CDM cosmology with $H_0=70$km s$^{-1}$ Mpc$^{-1}$, $\Omega_m=0.315$.

\begin{figure*}[th!]
    \centering
    \includegraphics[width=\textwidth,trim={0cm 3.25cm 14.3cm 0cm},clip]{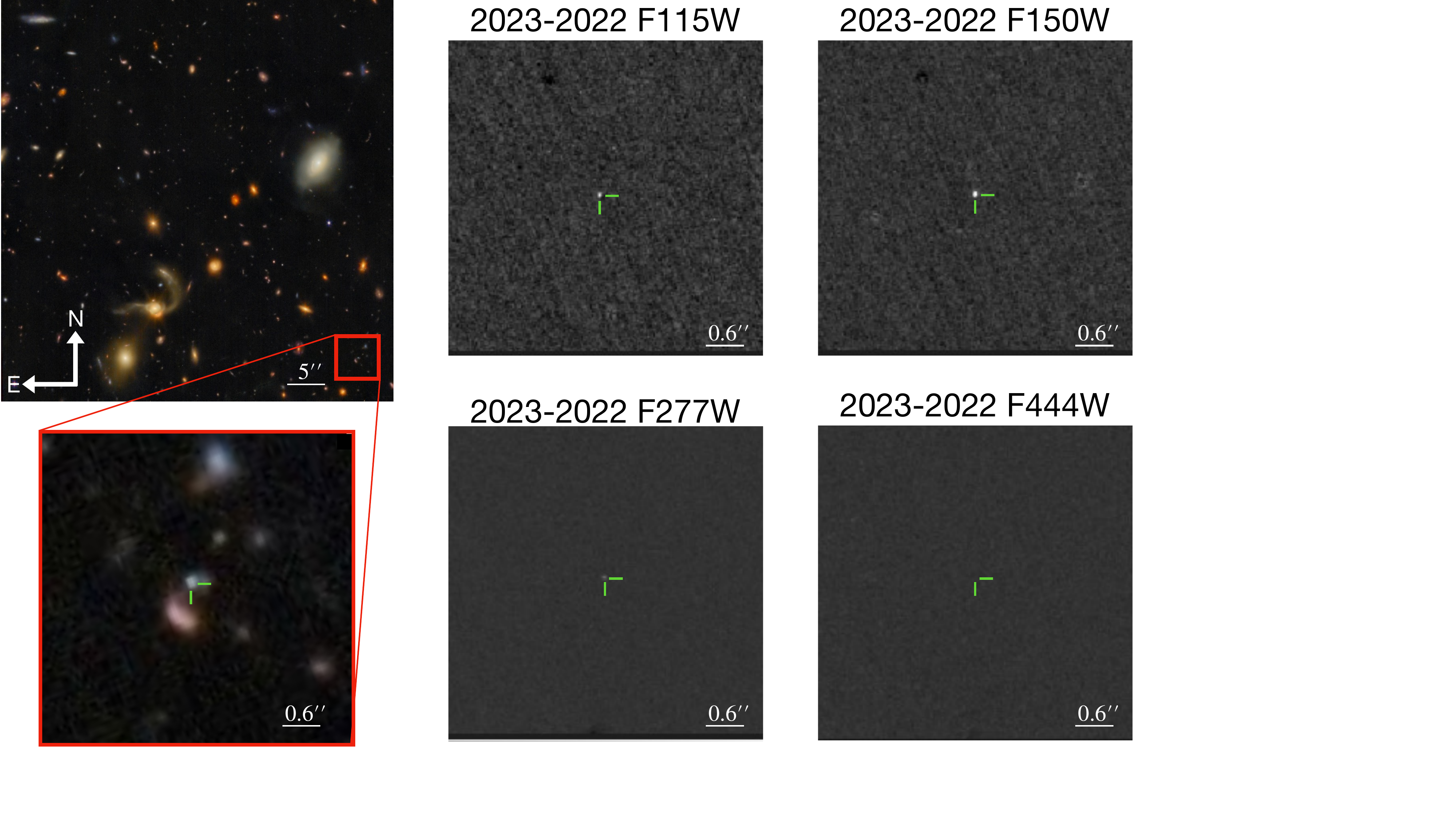}
    \caption{(Left column) Full-color images (from PID 2514) using F115W+F150W (Blue) F200W+F277W (Green) and F356W+F444W (Red), with a zoomed-out view on the top and the host of \highzIa on the bottom. (Column $2$-$3$) Difference images created from the COSMOS-Web and PRIMER visits separated by $\sim1$ year, with the \highzIa position marked with a green indicator. All images are drizzled to $0.03^{\prime\prime}/$pix and the cutouts have the same spatial extent. }
    \label{fig:im_cutouts}
\end{figure*}


\section{Summary of Observations}
\label{sec:obs}
\highzIa was observed in a total of three \textit{JWST} visits spanning $\sim200$ observer-frame days. The method for detecting SNe and the subsequent DDT program observations are described in detail by A.~Rest et al., in preparation (hereafter R24). Briefly, \highzIa was detected in COSMOS-Web observations (Program ID [PID] $1727$) taken on $2023$ December $27$ (MJD $\sim60305$) using the F115W, F150W, F277W, F444W filters. These overlapped with observations from the PRIMER program \citep{dunlop_primer_2021} taken in the observing window $2023$ May $6$-$18$ (MJD $\sim60076$), a baseline of $\sim2.5$ rest-frame months for \highzIa. A \textit{JWST} DDT program \citep[PID $6585$][]{coulter_high-z_2024} was approved to follow the most interesting transients with a visit on $2024$ April $29$ (MJD $60429$), including nearly seven hours of integration in the NIRSpec \citep{jakobsen_near-infrared_2022} multi-object spectroscopy (MOS) mode using the micro-shutter assembly \citep[MSA;][]{ferruit_near-infrared_2022} and low-resolution Prism (R$\sim100$). The MSA provided SN spectra for several transients, with the others described in companion papers (e.g., D.~Coulter et al. in preparation and R24) as well as a variety of galaxy spectra. 

After the discovery of \highzIa in COSMOS-Web data, an archival search of \textit{JWST}/NIRCam imaging revealed serendipitous observations by ``A Pure Parallel Wide Area Legacy Imaging Survey at $1$-$5$ Micron'' \citep[PANORAMIC, PID $2514$;][]{williams_panoramic_2021} on $2023$ December $3$ (MJD $60281$) in six NIRCam filters, preceding the COSMOS-Web detection image. PANORAMIC therefore provides the first epoch of imaging for \highzIa, and below we describe the data reduction and analysis of the combined dataset.

\subsection{Measuring Photometry}
\label{sub:obs_phot}
We follow the same methods for photometry on Level 3 (drizzled, I2D) \textit{JWST} images as \citet{pierel_discovery_2024}. Level 3 NIRCam images are the drizzled and resampled combination of Level 2 (CAL) NIRCam images. CALs are individual exposures that have been calibrated using the STScI JWST Pipeline\footnote{\url{https://github.com/spacetelescope/jwst}}
\citep{bushouse_jwst_2022},  
and have been bias-subtracted, dark-subtracted, and flat-fielded but not yet corrected for geometric distortion. 

We first align the individual NIRCam exposures containing \highzIa, from all three programs described above (see Table \ref{tab:phot}), to the drizzled template images from PRIMER, as PRIMER provides a temporal reference image in all filters where \highzIa was observed. We use the \textit{JWST}/\textit{HST} Alignment Tool \citep[{\tt JHAT};][]{rest_arminrestjhat_2023}\footnote{\url{https://jhat.readthedocs.io}}), which improves the relative default alignment from $\sim1$\,pixel to $\sim0.1$\,pixel between the epochs. 

We produce aligned drizzled images with the \textit{JWST} pipeline \citep[v$1.12.5$;][]{bushouse_jwst_2022}, and obtain difference images in all filters (Figure \ref{fig:im_cutouts}) using the High Order Transform of PSF and Template Subtraction \citep[{\tt HOTPANTS};][]{becker_hotpants_2015}\footnote{\url{https://github.com/acbecker/hotpants}}) code \citep[with modifications implemented in the \texttt{photpipe} code;][]{rest_testing_2005}. We implement the \texttt{space\_phot} \citep{pierel_space-phot_2024}\footnote{\url{space-phot.readthedocs.io}} drizzled PSF fitting routine using $5\times5$ pixel cutouts and Level 2 PSF models from {\tt webbpsf}\footnote{\url{https://webbpsf.readthedocs.io}}, which are temporally and spatially dependent and include a correction to account for the finite PSF size. The Level 2 PSF models are drizzled together using the same pipeline implementation as the data, and fit to the observed \highzIa flux. These total fluxes, which are in units of MJy/sr, are converted to AB magnitudes using the native pixel scale of each image ($0.03\arcsec/$pix for SW, $0.06\arcsec/$pix for LW). Measured photometry is given in Table \ref{tab:phot}. A final source of photometric uncertainty is a systematic uncertainty on the zero-points, which is $\lesssim0.01$ mag for all filters and is therefore subdominant to the uncertainties derived here (M. Boyer 2024, private communication; M. Boyer et al., in preparation).

\begin{table}
    \centering
    \caption{\label{tab:phot} Observations for \highzIa discussed in Section \ref{sec:obs}.}
    
    \begin{tabular*}{\linewidth}{@{\extracolsep{\stretch{1}}}*{5}{c}}
\toprule
PID&MJD&Instrument&\multicolumn{1}{c}{Filter/Disperser}&\multicolumn{1}{c}{m$_{AB}$}\\
\hline
$1837$&$60061$&NIRCam&F115W&--\\
$1837$&$60061$&NIRCam&F150W&--\\
$1837$&$60061$&NIRCam&F200W&--\\
$1837$&$60061$&NIRCam&F277W&--\\
$1837$&$60061$&NIRCam&F356W&--\\
$1837$&$60061$&NIRCam&F444W&--\\
\hline
$2514$&$60281$&NIRCam&F115W&$25.69\pm0.05$\\
$2514$&$60281$&NIRCam&F150W&$25.34\pm0.04$\\
$2514$&$60281$&NIRCam&F200W&$25.42\pm0.04$\\
$2514$&$60281$&NIRCam&F277W&$26.50\pm0.06$\\
$2514$&$60281$&NIRCam&F356W&$27.24\pm0.12$\\
$2514$&$60281$&NIRCam&F444W&$27.45\pm0.25$\\
\hline
$1727$&$60305$&NIRCam&F115W&$26.49\pm0.11$\\
$1727$&$60304$&NIRCam&F150W&$26.00\pm0.06$\\
$1727$&$60303$&NIRCam&F277W&$27.10\pm0.09$\\
$1727$&$60302$&NIRCam&F444W&$>28.5$\\
\hline
$6585$&$60429$&NIRCam&F115W&$28.97\pm0.26$\\
$6585$&$60429$&NIRCam&F150W&$28.08\pm0.09$\\
$6585$&$60429$&NIRCam&F277W&$28.34\pm0.19$\\
$6585$&$60429$&NIRCam&F444W&$>28.6$\\
\hline
$6585$&$60429$&NIRSpec&Prism&--\\
\hline
\hline
    \end{tabular*}
\begin{flushleft}
\tablecomments{Columns are: \textit{JWST} Program ID, Modified Julian date, \textit{JWST} instrument, filter or grating, and photometry plus final uncertainty for \highzIa. Upper limits are $5\sigma$. PID $1837$ is PRIMER, $2514$ is PANORAMIC, $1727$ is COSMOS-Web, and $6585$ is the DDT program. Each magnitude is measured using a difference image with PRIMER (PID $1837$).}

\end{flushleft}
\end{table}


\subsection{NIRSpec Reduction}
\label{sub:obs_spec}

\begin{figure*}
    \centering
    \includegraphics[width=\textwidth,trim={7cm 0cm 5cm 0cm},clip]{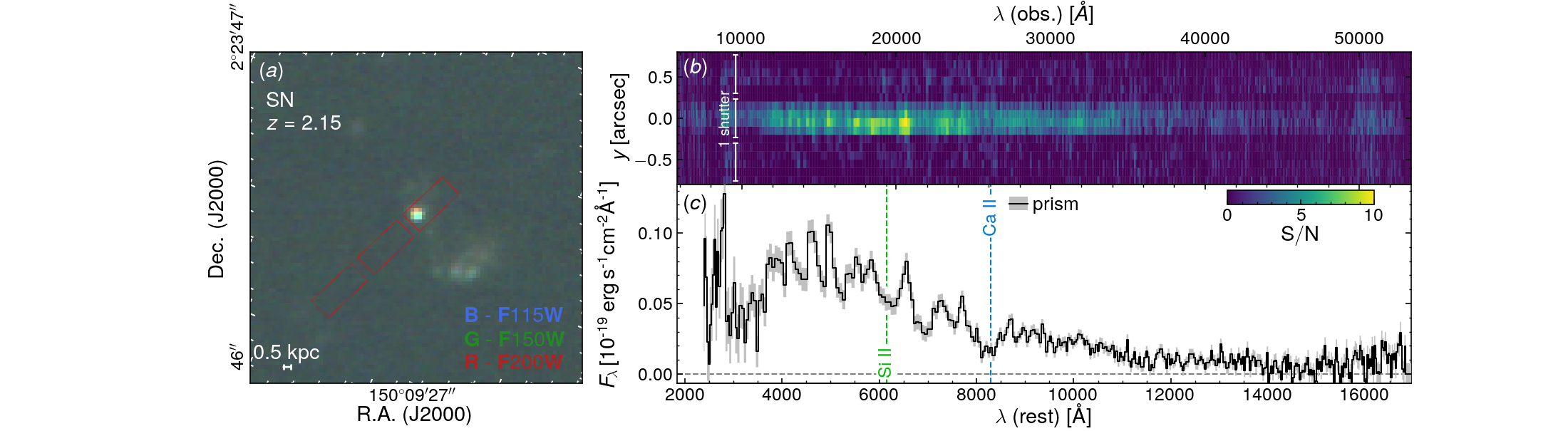}
    \caption{(a) The NIRSpec/MSA slitlet positions over \highzIa for one of the dithered observations. (b,c) The 2D and 1D-extracted NIRSpec spectrum for \highzIa. The two absorption lines primarily used for spectroscopic classification of SNe\,Ia, are shown with dotted lines.  }
    \label{fig:spec_2d_cuts}
\end{figure*}

\begin{figure}
    \centering
    \includegraphics[trim={.5cm .5cm 2cm 1cm}, clip,width=\linewidth]{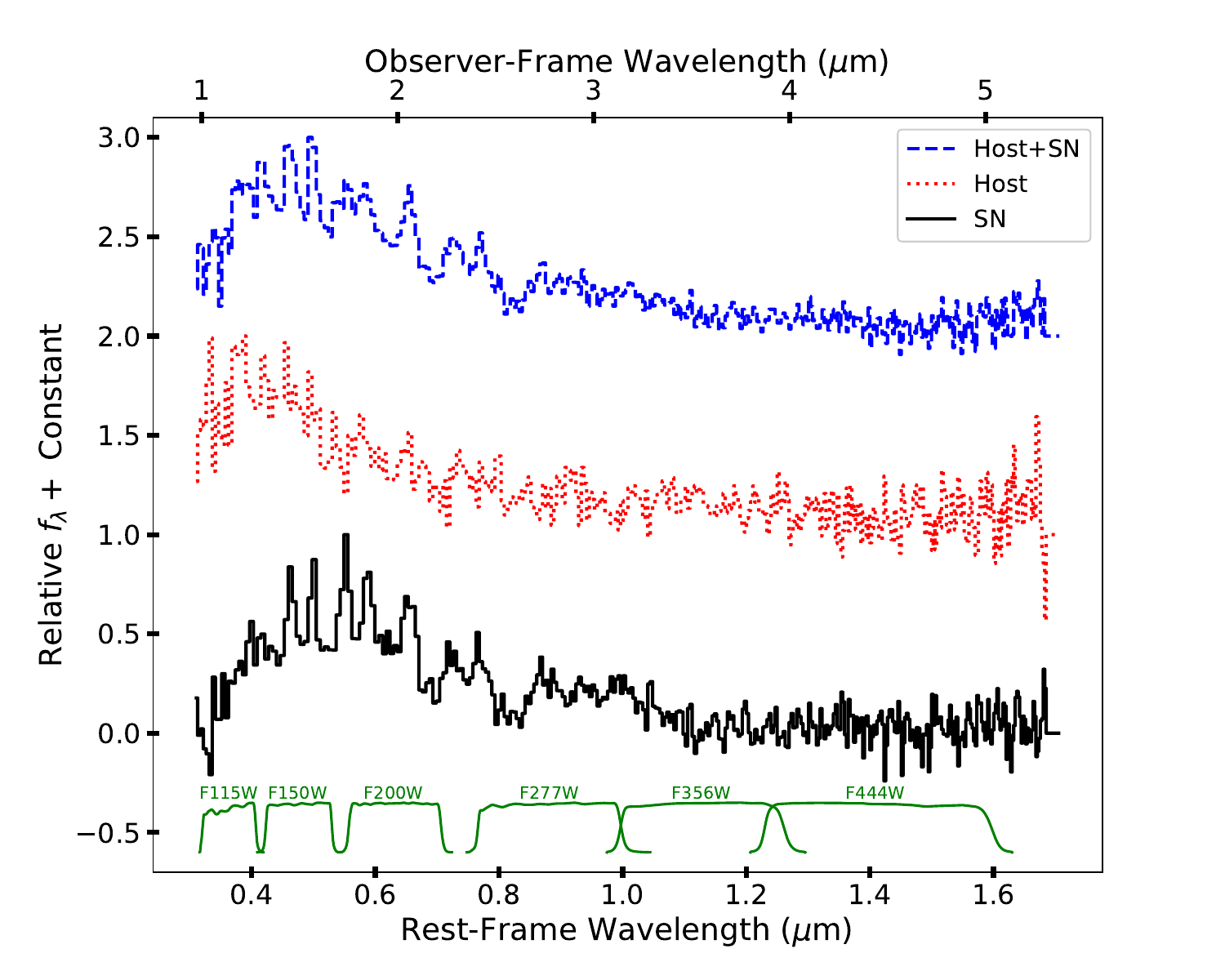}
    \caption{The raw combined Host$+$SN spectrum is shown as a blue dashed line (top), followed by the host spectrum as a red dotted line extracted from rows neighboring the SN trace in Figure \ref{fig:spec_2d_cuts} (middle), and the host-subtracted SN spectrum is a black solid line (bottom). The black line is used for spectroscopic classification, and no obvious spectroscopic emission features are present for a redshift in the host spectrum. The NIRCam filters used here are shown in green at the bottom of the figure for reference.}
    \label{fig:spec_decomp}
\end{figure}

We began processing the spectroscopic data with Stage 2 products from the Mikulski Archive for Space Telescopes (MAST). Additional processing used the \textit{JWST} pipeline \citep[v$1.12.5$;][]{bushouse_jwst_2022} \footnote{With context file jwst\_1183.pmap} to produce two-dimensional (2D) spectral data (Figure \ref{fig:spec_2d_cuts}). The pipeline applied a slit-loss throughput correction for \highzIa based on the planned position of a point-source within the MSA shutters (Figure \ref{fig:spec_2d_cuts}). The 2D spectrum of \highzIa and its host galaxy was extracted using the
optimal extraction algorithm from \citet{horne_optimal_1986} implemented as scripts available as part of the MOS Optimal Spectral Extraction (MOSE) notebook\footnote{\url{https://spacetelescope.github.io/jdat_notebooks/notebooks/ifu_optimal/ifu_optimal.html}}. To separate the spectrum of \highzIa and its host galaxy we extract a 1D spectrum from each of the rows neighboring the trace associated with \highzIa, which creates a spectrum of the host galaxy from the region local to the SN. We then subtract this host galaxy spectrum from the combined extraction to obtain a 1D spectrum for \highzIa, with each of these stages shown in Figure \ref{fig:spec_decomp}. We discuss the effectiveness of this method in the next section.

\section{Classification as Type Ia}
\label{sec:class}
There are no clear $[\rm{O}III]$ (rest-frame $5,007\angstrom$) or $\rm{H}\alpha$ ($6,563\angstrom$) emission lines in the host galaxy spectrum (Figure \ref{fig:spec_decomp}), making a precise spectroscopic redshift measurement impossible. We therefore first fit the template host photometry, which was measured from observations taken in $2022$ by the PRIMER program \citep{dunlop_primer_2021} as well as earlier optical \textit{HST} imaging from COSMOS \citep{koekemoer_cosmos_2007} and CANDELS \citep{grogin_candels_2011,koekemoer_candels_2011}, and \textit{Spitzer} mid-IR imaging, with the EAZY software package \citep{brammer_eazy_2008}. The SED was well sampled, and provided a robust photo-$z$ posterior with a $68^{\rm{th}}$-percentile range of $1.91<z<2.29$, with a best-fit of $z=2.25$. Other basic properties of the host from the SED fit are given in Table \ref{tab:host}, derived after fixing the redshift based on the analysis below. The properties show a relatively low-mass quiescent galaxy, which is expected for an SN\,Ia.

\begin{table}
    \centering
    \caption{\label{tab:host} Properties derived from an EAZY fit to the host SED. The redshift was measured before follow-up, the other properties after fixing the redshift to $z=2.15$ given this analysis.}
    
    \begin{tabular*}{\linewidth}{@{\extracolsep{\stretch{1}}}*{4}{c}}
\toprule
&Property&Best-fit&\\
\hline
&$z$&$2.25^{+0.04}_{-0.34}$&\\
\hline
&$\rm{log} M_*/M_\odot$&$8.1\pm0.1$&\\
&$\rm{log} Z_*/Z_\odot$&$-1.0\pm0.3$&\\
&$\rm{log} sSFR$&$-8.2\pm0.2$&\\
&$A_V$&$0.2^{+0.2}_{-0.1}$&\\
\end{tabular*}
\begin{flushleft}
\tablecomments{Columns are: Host property and best-fit measurement with uncertainty. The measured properties are redshift, mass, metallicity, specific star formation rate, and V-band dust extinction. The redshift is that measured before follow-up using the PRIMER SED (setting the prior for this analysis), while the remaining properties are derived after setting the redshift to $z=2.15$ based on this analysis.}

\end{flushleft}
\end{table}

We use the Next Generation SuperFit \citep[\texttt{NGSF};][]{goldwasser_next_2022}\footnote{\href{https://github.com/oyaron/NGSF}{https://github.com/oyaron/NGSF}} package to classify \highzIa. NGSF matches a database of reference SN spectra for a variety of SN types to the observed spectrum, while varying levels of dust extinction and host galaxy contamination. We allow the redshift to vary in the $1\sigma$ range of the photo-$z$ for the host galaxy ($1.9<z<2.3$). Of the top ten reference SN spectra matched to the \highzIa spectrum, eight are of normal SNe\,Ia and the remainder are the $91$T SN\,Ia sub-type \citep[e.g.,][]{fisher_spectrum_1999}, with the best match being the normal SN\,Ia $2011$by at $+52$ rest-frame days relative to peak B-band brightness \citep[e.g.,][]{foley_classifying_2013,graham_twins_2015,foley_significant_2020}. A variety of normal and peculiar SN\,Ia sub-types fill the top $40$ matched reference spectra, before core-collapse subtypes appear. 

\begin{figure*}[t!]
    \centering
    \includegraphics[width=\textwidth,trim={1cm 0cm 1.25cm -2cm},clip]{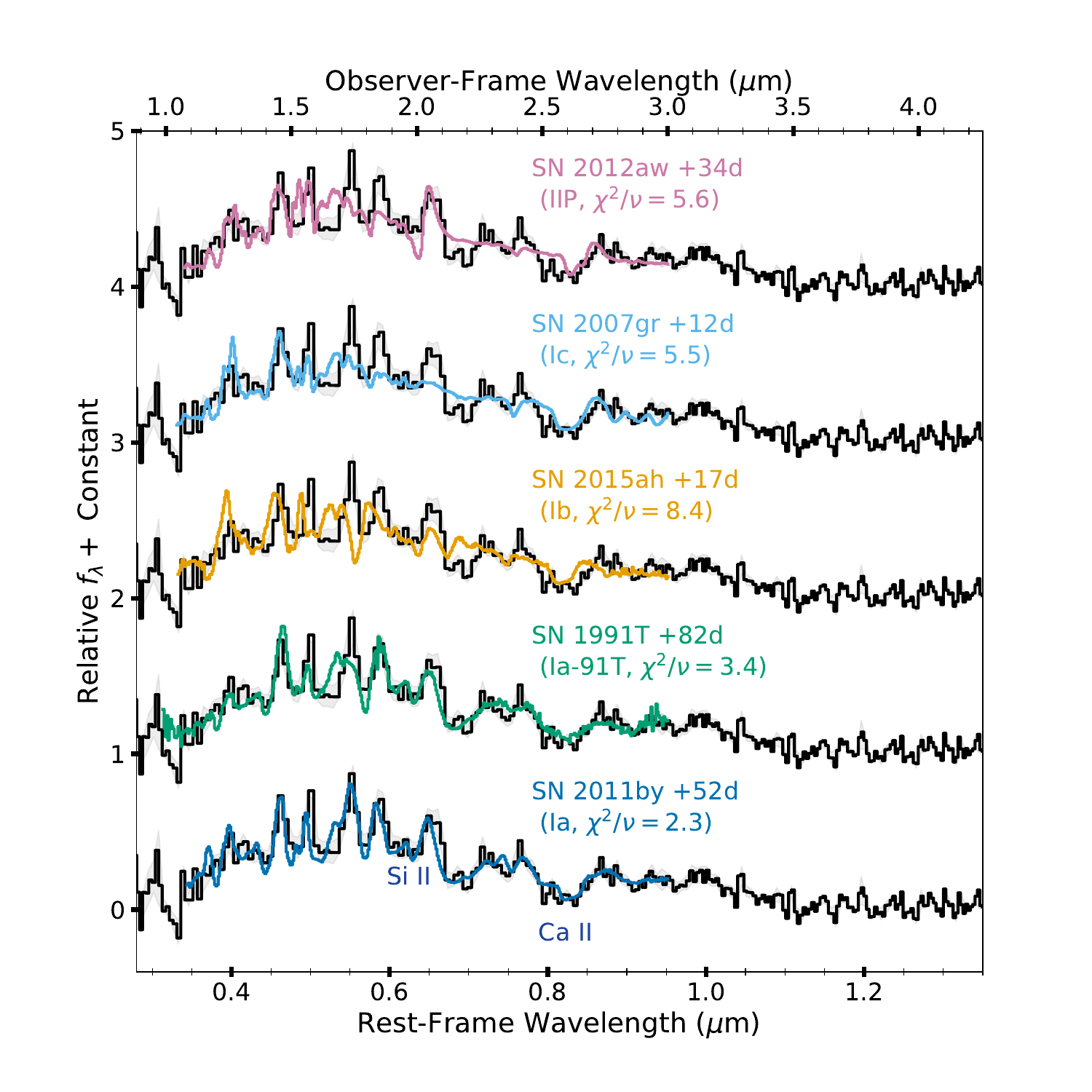}
        \caption{The host-subtracted NIRSpec spectrum (with uncertainty) of \highzIa is shown as a black solid line, with the primary features used for the preferred SN\,Ia classification labeled (bottom). The best-match template from \texttt{NGSF} is a SN\,Ia (blue,$+52d$, bottom), and a $91T$-like SN\,Ia subclass (green, $+82d$, second from bottom)  is also shown for comparison. The best core-collapse matches are also shown, including Ib (yellow, third from top), Ic (cyan, second from top), and IIP (pink, top). A normal SN\,Ia is heavily favored based on the spectrum (see Table \ref{tab:phase}). All phases are rest-frame days relative to peak B-band brightness.}
    \label{fig:sn_spec}
\end{figure*}

Figure \ref{fig:sn_spec} shows the best spectral template match for SN\,Ia, SN\,Ia $91$T-like, and three core-collapse sub-types (IIP, Ib, Ic). All templates show a $8300\angstrom$ Ca\,II absorption, but the shape of the absorption is best-matched by the SN\,Ia sub-types. The $6150\angstrom$ Si\,II feature is also present in both SN\,Ia sub-type templates as well as the \highzIa spectrum, but not the core-collapse sub-types. Additionally, there are a variety of features present in the various core-collapse sub-types not present in the \highzIa spectrum. Overall, the SN\,Ia spectral template match provides a $\chi^2$ per degree of freedom ($\nu$) of $2.30$, while the next best (non-normal SN\,Ia) fit is a $91$T-like SN\,Ia with $3.40$ (Table \ref{tab:phase}). 

Table \ref{tab:phase} also contains the inferred time of peak B-band brightness for \highzIa, based on the phase of the best-fit spectral template for each SN sub-type. From the light curve fitting in Section \ref{sec:distance}, it is clear that the time of peak brightness must be before MJD $60305$ (as the epoch at $60281$ is brighter), a condition that is only met by the SN\,Ia and $91$T-like templates. In other words, if \highzIa would be a core-collapse SN, then based on the spectrum it would need to be younger than what is possible given the light curve (Section \ref{sec:distance}). Of the two SN\,Ia classes, a normal SN\,Ia is strongly preferred given the significantly lower $\chi^2/\nu$ (Table \ref{tab:phase}).  Additionally, the light curve decline rate is more consistent with a normal SN\,Ia \citep[see Section \ref{sub:distance_lc};][]{dimitriadis_ztf_2024}, so we therefore proceed with the classification of normal SN\,Ia. 

We take the median and standard deviation in redshift for the top ten SN\,Ia spectral matches as the final redshift measurement for \highzIa, which is $\snz\pm\snzerr$. We also note that NGSF provides an estimate of the fraction of flux in the spectrum associated with the SN relative to the host galaxy. Running NGSF on the blue spectrum in Figure \ref{fig:spec_decomp} decreases the SN flux contribution of the top matches to $\sim40\%$ from $\gtrsim90\%$ (for the black spectrum) while improving the $\chi^2/\nu$, giving some evidence of the effectiveness of the method.  

\begin{table}
    \centering
    \caption{\label{tab:phase} The time of peak B-band brightness ($t_{pk}$) inferred from the best spectral template match for each SN type, and the resulting reduced $\chi^2$. }
    
    \begin{tabular*}{\linewidth}{@{\extracolsep{\stretch{1}}}*{3}{c}}
\toprule
SN Type&Best $t_{pk}$&$\chi^2/\nu$\\
\hline
\textbf{Ia}&$\mathbf{60265}$&$\mathbf{2.30}$\\
Ia-91T&$60171$&$3.40$\\
Ib&$60374$&$8.37$\\
Ic&$60391$&$5.52$\\
IIP&$60321$&$5.57$\\

\end{tabular*}
\begin{flushleft}
\tablecomments{Columns are: SN type model/spectral template used, the time of peak B-band brightness given the best-fit spectroscopic template match, and the $\chi^2$ per DOF of the best-fit spectroscopic template match. Note that given the temporal separation between detection and obtaining the spectrum of \highzIa the time of peak brightness must be MJD$<60305$, which is only satisfied by SN\,Ia and Ia-91T with a statistical preference for normal SN\,Ia.}

\end{flushleft}
\end{table}

\begin{figure}
    \centering
    \includegraphics[width=.5\textwidth,trim={0.25cm 0.5cm 1cm 2.5cm},clip]{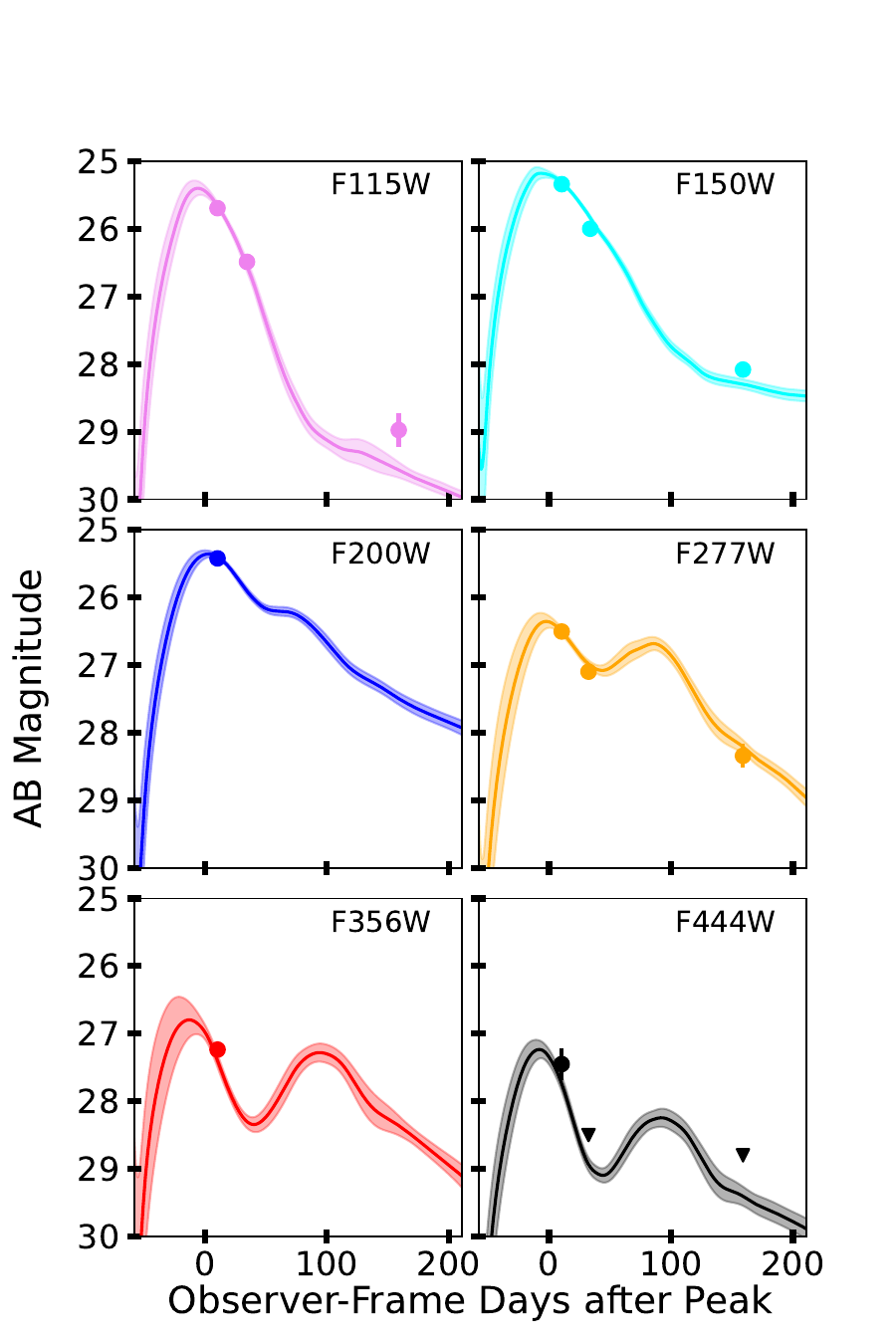}
    \caption{The photometry measured in Section \ref{sub:obs_phot} is shown as black circles with error, with ($5\sigma$) upper-limits denoted by triangles. The fit to the full light curve is shown as a solid line in each filter, with the shaded region representing the BayeSN model$+$fitting uncertainty. The time of peak brightness is measured relative to the rest-frame B-band and is reported in Table \ref{tab:fit}, and the amplitude is set by the luminosity distance (Section \ref{sub:distance_lc}). }
    \label{fig:lc_fit}
\end{figure}

\begin{figure}
    \centering
    \includegraphics[width=.5\textwidth,trim={0.25cm 0.25cm 1cm 1cm},clip]{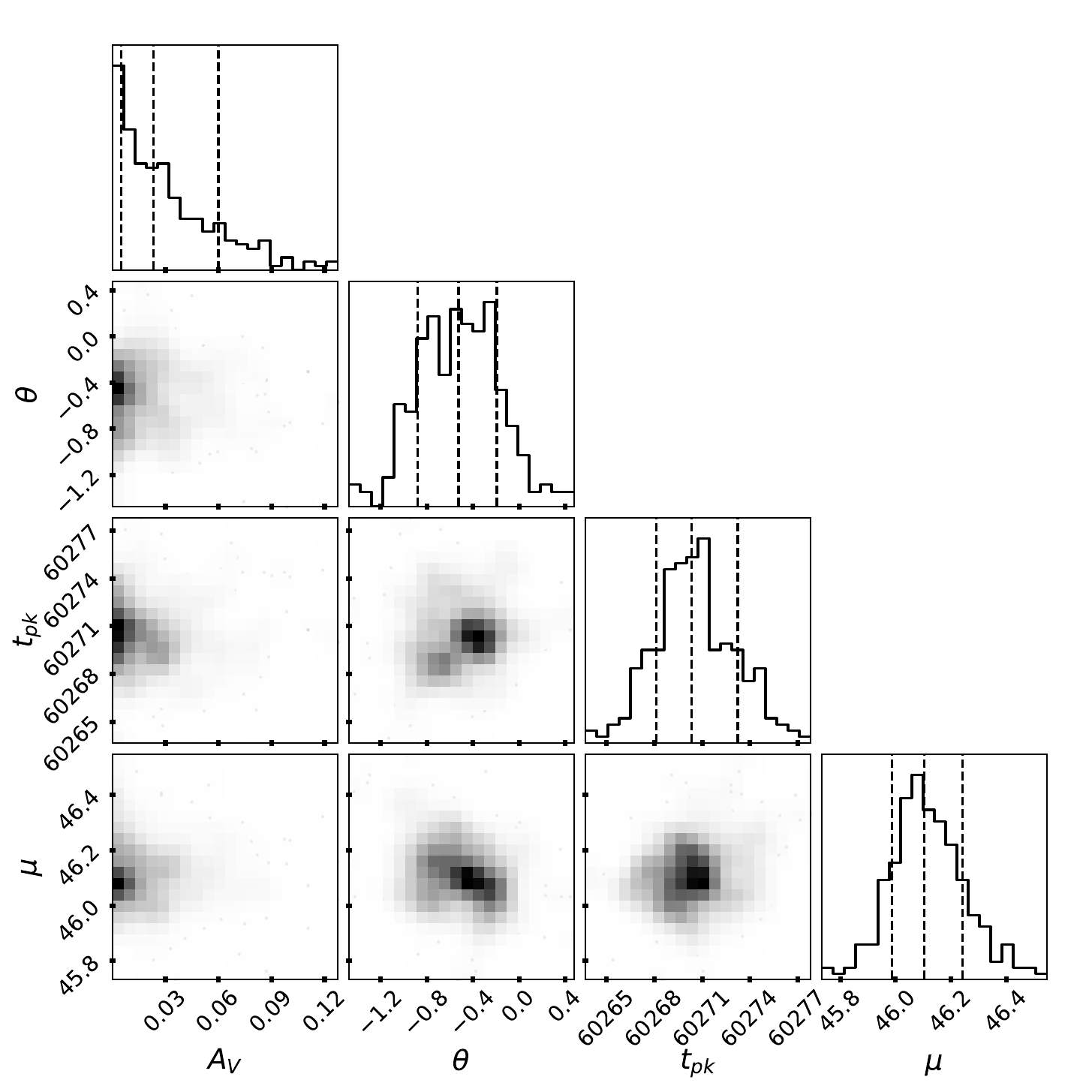}
    \caption{Posterior distributions for the BayeSN fitting of \highzIa photometry. The dashed vertical lines correspond to the distribution $16^{\rm{th}} , \ 50^{\rm{th}}, \ \rm{and} \ 84^{\rm{th}}$ quantiles. The parameters are $V$-band dust extinction, light curve shape, time of peak B-band brightness, and distance modulus. }
    \label{fig:fit_post}
\end{figure}

\section{Luminosity Distance Measurement}
\label{sec:distance}

\subsection{Light Curve Fitting}
\label{sub:distance_lc}

\begin{table}
    \centering
    \caption{\label{tab:fit} The Bayesn light curve model parameters used in this analysis.}
    
    \begin{tabular*}{.8\linewidth}{@{\extracolsep{\stretch{1}}}*{2}{c}}
\toprule
Parameter&Best-Fit\\
\hline
$z$&$\snz$ (Fixed)\\
$t_{pk}$ (MJD)&$60270.34^{+2.89}_{-2.23}$\\
$A_V$&$<0.04$\\
$\theta$&$-0.52^{+0.33}_{-0.36}$\\
$\mu$&$46.10^{+0.14}_{-0.12}$\\
\end{tabular*}
\begin{flushleft}
\tablecomments{Columns are: BayeSN parameter, and fitted result with uncertainty. Note that the fitted $A_V$ is approaching the physical lower bound of $0$, and so we report the $1\sigma$ upper-limit. The uncertainty on $\mu$ includes only model and statistical uncertainties, with more added in Section \ref{sec:distance}. The best-fit MJD corresponds to November $22$, $2023$.}
\end{flushleft}
\end{table}
We begin by fitting the observed photometry (including upper-limits; Table \ref{tab:phot}) with a phase-extended version of the BayeSN SN\,Ia SED model \citep{mandel_hierarchical_2022,grayling_scalable_2024}, which covers the full plausible rest-frame phase range of \highzIa, to fit the measured photometry. The version of BayeSN we use is a variant of the one presented by \citet{ward_relative_2023} and uses the training set detailed therein. The model is trained to cover rest-frame phases as late as $50$ days after peak-brightness with linear extrapolation in the space of the logarithm of the SED beyond that. This is the same model used by \citet{pierel_jwst_2024} to successfully fit a (strongly lensed) SN\,Ia at $z=1.78$, and is the only current light curve model with the phase and wavelength range needed to fit these data.

\begin{figure}
    \centering
    \includegraphics[width=.38\textwidth,trim={.25cm .25cm .25cm 0cm},clip]{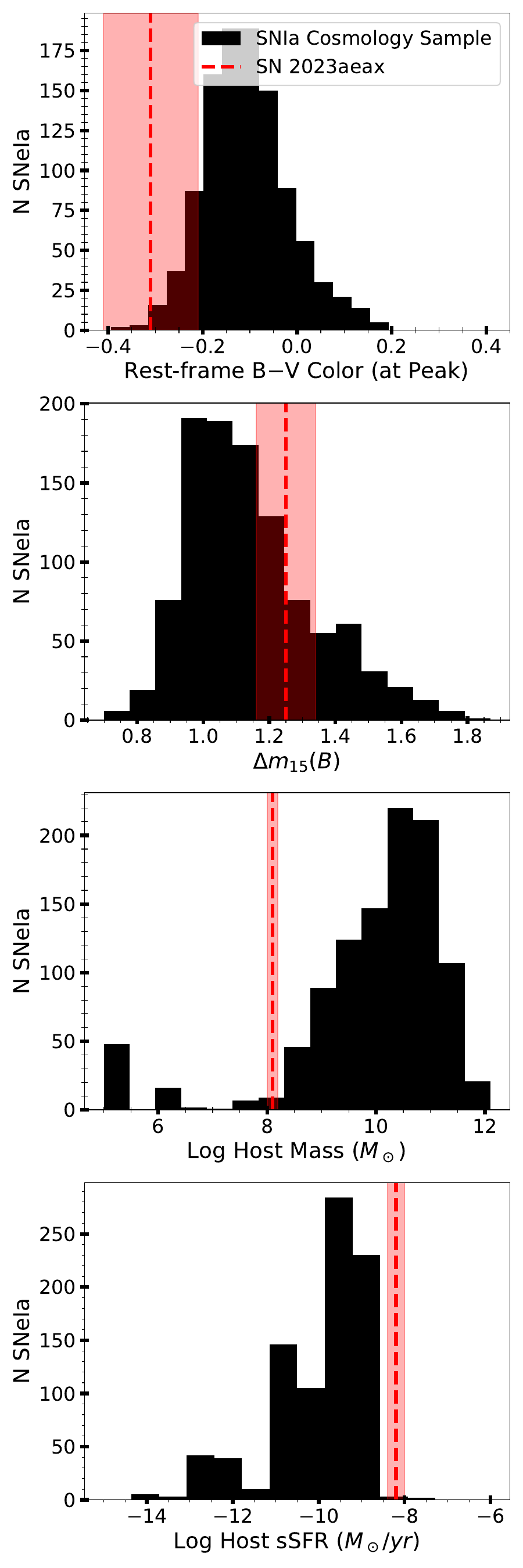}
    \caption{The distributions of rest-frame B$-$V color at peak B-band brightness, $\Delta m_{15}(B)$, and host galaxy properties measured for the cosmological sample of SNe\,Ia presented in \citet{brout_pantheon_2022}. The vertical red dashed lines are the measurements for \highzIa.}
    \label{fig:color_dist}
\end{figure}

In addition to the base template, BayeSN includes a rest-frame host-galaxy dust component (parameterized by the $V$-band extinction $A_V$ and ratio of total to selective extinction $R_V$) and a light curve shape parameter $\theta$ (This has a strong negative correlation with the SALT2 light curve model stretch parameter, $x_1$, meaning positive $\theta$ implies a faster decline rate; see e.g., Figure 4 of \citealp{mandel_hierarchical_2022}). We add a $0.01$\,mag Galactic extinction correction to the model based on the dust maps of \citet{schlafly_measuring_2011} and using the extinction curve from \citet{fitzpatrick_correcting_1999}. The BayeSN ``$\epsilon$-surface'' captures all additional model/intrinsic scatter and intrinsic color variation. Finally, BayeSN directly infers the luminosity distances as part of the Bayesian inference process. The best-fit model is shown with the observed photometry in Figure \ref{fig:lc_fit}, and the retrieved BayeSN parameters from the fit are shown in Table \ref{tab:fit}, with the posterior distributions shown in Figure \ref{fig:fit_post}. Note that the fit for $A_V$, the extrinsic dust, approaches the physical lower bound of $0$. This suggests little to no dust extinction for \highzIa (in agreement with the host SED fitting in Table \ref{tab:host}), but as this is not a reliable constraint on $A_V$ we report the $1\sigma$ upper-limit ($0.04$). The value of $\theta$ corresponds to a $\Delta m_{15}(B)$ of $\sim1.25\pm0.09$, which is well-within the normal low-$z$ SN\,Ia population (Figure \ref{fig:color_dist}. The rest-frame $B-V$ color at peak brightness for \highzIa is $-0.3\pm0.1$mag, which is on the blue edge of the normal low-$z$ SN population (see Figure \ref{fig:color_dist}), but is just within traditional low-$z$ color cosmology cuts \citep[$-0.4\lesssim B-V\lesssim0.4$, or using the SALT ``$c$'' parameter $-0.3<c<0.3$;][]{brout_pantheon_2022}.

\subsection{Adding \highzIa to the Hubble Diagram}
\label{sub:distance_hd}
 As noted in the previous section, BayeSN infers the luminosity distance for a SN\,Ia directly as a result of its Bayesian model fitting step. The only additional corrections to this value would be a bias correction \citep[e.g.,][]{kessler_correcting_2017} and a correction for the host-galaxy mass step, or the small residual correlation between SN\,Ia distance measurements and their host-galaxy masses that may be driven by interstellar medium (ISM) dust mass surface density \citep{kelly_hubble_2010,lampeitl_effect_2010,sullivan_dependence_2010,brout_its_2021}. The nature and evolution of the host-galaxy mass step is unknown, especially at such high redshift \citep[e.g.,][]{childress_ages_2014}, and its correction to a BayeSN distance measurement is not well-understood. We simply apply half of the host mass step (for a low-mass galaxy, Table \ref{tab:host}) from \citet{brout_pantheon_2022} (who found $\sim0.054$mag) as a systematic error in quadrature. As for the selection bias correction, \citet{pierel_discovery_2024} explored this with extensive simulations for an SN\,Ia also discovered at $z>2$ by \textit{JWST}. \highzIa is several magnitudes brighter than the survey limiting magnitude, and in this case the fitted light curve parameters are within normal ranges,  meaning that a bias correction is not necessary in this instance. Once a larger sample of $z>2$ SNe\,Ia are found, a population-level analysis will be warranted to account for biases \citep[e.g.,][]{scolnic_complete_2018,brout_pantheon_2022}.

\begin{figure*}[h!]
    \centering
    \includegraphics[width=.97\textwidth,trim={.5cm 0cm 2.5cm 0cm},clip]{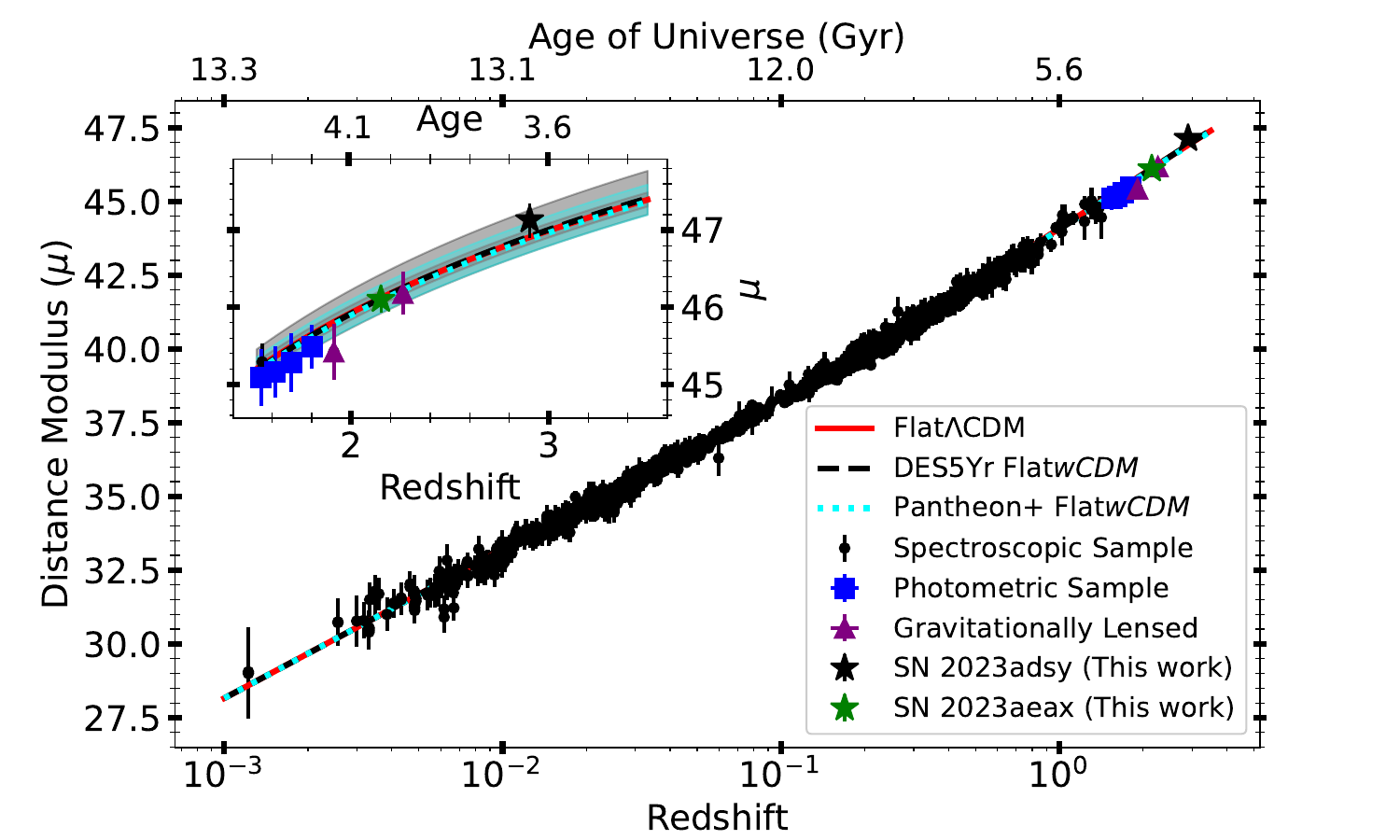}
    \caption{Luminosity distance measurements from the full sample of SNe\,Ia from \citet{brout_pantheon_2022} extending to $z=2.22$. Black points (with errors) are SNe\,Ia with spectroscopic classifications, while blue squares (with error) are SNe\,Ia with photometric classifications. The two strongly lensed SNe\,Ia with distance measurements are shown as purple triangles. \highzIa is shown as a green star, and $\Lambda$CDM is shown as a solid red line. The width of the red line encompasses the width of the current $H_0$ tension, with the center of the line used for reference. The black star is the distance modulus for SN\,$2023$adsy measured in this work with photometry from \citet{pierel_discovery_2024}. The best-fit Flat$wCDM$ cosmological constraints from \citet{brout_pantheon_2022} (black-dashed) and \citet{des_collaboration_dark_2024} (cyan-dotted) are also shown for comparison.}
    \label{fig:lcdm}
\end{figure*}
The final luminosity distance measurement is $\mufit_{-\mufiterrl}^{+\mufiterrh}$\,mag, while the $\Lambda$CDM prediction at $z=\snz$ (with $H_0=70$ km s$^{-1}$ Mpc$^{-1}$) is $\mu=46.12$\,mag, a $\ll1\sigma$ difference (Figure \ref{fig:lcdm}). The uncertainty on $\mu$ includes the fitted model uncertainties, errors from redshift and peculiar velocity (which are negligible here), and an additional $0.055z$\,mag uncertainty from weak gravitational lensing \citep{jonsson_constraining_2010}. \highzIa is only the second spectroscopically classified non-strongly lensed SN\,Ia at $z>1.6$, and while its distance modulus agrees with $\Lambda$CDM the combination of $z<2$ SNe\,Ia and SN\,$2023$adsy with \highzIa suggests a possible slope (at $\sim1\sigma$ relative to $\Lambda$CDM) in SN\,Ia luminosity distance measurements as a function of redshift. This trend is well-within the best current cosmological constraints from SNe\,Ia \citep{brout_pantheon_2022,des_collaboration_dark_2024}, which are shown in Figure \ref{fig:lcdm}.  We note that the SN\,$2023$adsy luminosity distance was measured with SALT3-NIR \citep{pierel_salt3nir_2022}, but we re-fit the published photometry (including the $>4\mu$m data) following our methodology here and find excellent agreement between the two models ( $47.14_{-0.24}^{+0.21}$ compared to $47.18_{-0.28}^{+0.27}$). In Figure \ref{fig:lcdm}, we show the SN\,$2023$adsy distance modulus measured in this work alongside \highzIa for reference. 

\section{Discussion}
\label{sec:conclusion}
We have presented \textit{JWST} observations of a SN (\highzIa) with a redshift of $z=\snz\pm\snzerr$, which we classify using a NIRSpec spectrum as a SN\,Ia.  Using the BayeSN SN\,Ia standardization model that has been used successfully at lower redshift ($z\lesssim2$), we measure the luminosity distance to \highzIa. We find a value of $\mu=\mufit_{-\mufiterrl}^{+\mufiterrh}$\,mag, which is in $0.1\sigma$ agreement  with $\Lambda$CDM. Although more objects are needed to directly constrain cosmological parameters at high-$z$, any significant deviation of SN\,Ia luminosity distances from $\Lambda$CDM at $z>2$ (in the dark matter-dominated universe) would be a strong indicator of SN\,Ia luminosity evolution with redshift or evidence for a more exotic cosmological model. Such a test is only newly possible with \textit{JWST}, the only telescope capable of robust spectroscopy of SNe\,Ia in the dark matter-dominated universe. While \highzIa is in excellent agreement with $\Lambda$CDM, combining this SN with previous $z<2$ SNe\,Ia and SN\,$2023$adsy at $z=2.903$ results in a $\sim1\sigma$ slope relative to $\Lambda$CDM but well within current constraints from \citet{brout_pantheon_2022} and \citet{des_collaboration_dark_2024}, and so a larger sample needed to determine the validity of this result.

\highzIa is just the second spectroscopically confirmed, non-gravitationally lensed SN\,Ia at $z>2$ (the other is SN $2023$adsy at $z=2.903$), and the first with light curve properties that would pass low-$z$ cosmology cuts. The phase of our NIRSpec spectrum does not allow a direct comparison of spectral features between \highzIa and SN $2023$adsy, but we can compare light curve properties. We have fit both SNe\,Ia with BayeSN for direct comparison, removing the possibility of a model bias, resulting in a small slope from $z<2$ to $z\sim3$ at low confidence ($\sim1\sigma$). The SNe have normal light curve decline rates ($\Delta\rm{m}_{15}(B)=1.25\pm0.09$ for \highzIa), but SN $2023$adsy had a peculiarly red color (rest-frame $B-V$ of $\sim0.8$) that was not easily attributable to dust attenuation, compared to the peak rest-frame $B-V$ color for \highzIa of $-0.3\pm0.1$. We require a larger population of high-$z$ SNe\,Ia to determine if there are significant changes to the SN\,Ia population properties between $z=\snz$ and $z=2.9$, or if SN $2023$adsy is simply an outlier within the high-$z$ SN\,Ia population. While the peak color of \highzIa is considered just within the cosmological sample for low-$z$ SNe\,Ia, it is on the extreme blue end, giving another point of consideration as the population of SNe\,Ia at $z>2$ grows. The host galaxy properties of \highzIa are also on the edges of, but within, the existing sample of SNe\,Ia. It is unclear exactly how these properties will evolve with redshift, and such a result is not unexpected.

\highzIa is the second SN\,Ia with a combined spectroscopic and photometric dataset in the dark matter-dominated universe at $z>2$, and is the first test for standardized luminosity evolution using a SN\,Ia that would pass low-$z$ cosmology cuts. Now with a ``sample'' of two such objects, we find small peculiarities with both that could be indicative of interesting population changes with redshift, or simply statistical fluctuations. The combination of these two SNe\,Ia from \textit{JWST} with previous work at $z\lesssim2$ hints at a possible slope in distances with redshift that could be attributed to evolution, with more objects needed to confirm or refute this result. \textit{JWST} is expected to add $\gtrsim10$ additional such objects over the next two years \citep{pierel_pass_2024}, and is the only resource capable of doing so before the launch of the \textit{Nancy Grace Roman Space Telescope} \citep{hounsell_simulations_2018,rose_reference_2021}. Regardless, \textit{JWST} will remain the only telescope capable of $z>2$ SN\,Ia spectroscopy, and the sample of $\gtrsim10$ will be required to put valuable constraints on any possible evolution at lower redshift for future cosmological measurements.

\begin{center}
    \textbf{Acknowledgements}
\end{center}

We would like to thank Matthew Grayling for useful discussion. This paper is based in part on observations with the NASA/ESA Hubble Space Telescope and James Webb Space Telescope obtained from the Mikulski Archive for Space Telescopes at STScI. We thank the DDT and JWST/HST scheduling teams at STScI for extraordinary effort in getting the DDT observations used here scheduled quickly. This work is based on observations made with the NASA/ESA/CSA James Webb Space Telescope. The data were obtained from the Mikulski Archive for Space Telescopes at the Space Telescope Science Institute, which is operated by the Association of Universities for Research in Astronomy, Inc., under NASA contract NAS 5-03127 for JWST. These observations are associated with programs \#1727, 1837, 2514, and 6585.
This research is based (in part) on observations made with the NASA/ESA Hubble Space Telescope obtained from the Space Telescope Science Institute, which is operated by the Association of Universities for Research in Astronomy, Inc., under NASA contract NAS 5–26555. The JWST data used in this paper can be found in MAST: \dataset[10.17909/n7kq-ef83]{https://dx.doi.org/10.17909/n7kq-ef83}. 
JDRP is supported by NASA through an Einstein
Fellowship grant No. HF2-51541.001 awarded by the Space
Telescope Science Institute (STScI), which is operated by the
Association of Universities for Research in Astronomy, Inc.,
for NASA, under contract NAS5-26555. Support was provided to DC, AR, and ME through program HST-GO-6541. The Cosmic Dawn Center (DAWN) is funded by the Danish National Research Foundation (DNRF140). This work was made possible by utilising the CANDIDE cluster at the Institut
d’Astrophysique de Paris. The cluster was funded through grants
from the PNCG, CNES, DIM-ACAV, the Euclid Consortium, and
the Danish National Research Foundation Cosmic Dawn Center
(DNRF140). It is maintained by Stephane Rouberol.
M.R.S. is supported by the STScI postdoctoral fellowship. ZGL is supported by the Marsden Fund administered by the Royal Society of New Zealand, Te Apārangi under grant M1255.
JS is supported by JSPS KAKENHI (JP22H01262) and
the World Premier International Research Center Initiative (WPI), MEXT, Japan. MF is supported by the European Union’s Horizon 2020 research and innovation programme under the Marie Sklodowska-Curie grant agreement No 101148925.

\clearpage

\bibliographystyle{aasjournal}



\end{document}